\documentclass[sort&compress,10pt,twocolumn,final,3p]{elsarticle}

\usepackage[utf8]{inputenc}
\usepackage{hyperref}
\usepackage{graphicx}
\usepackage{subfigure}
\usepackage{array}
\usepackage{multirow}
\usepackage[]{footmisc}	
\usepackage{amsmath,amssymb}
\usepackage{bm}	
\usepackage{xspace} 

\usepackage{miller}

\newcommand{\etal}{\textit{et al}.\@\xspace}
\newcommand{\ie}{\textit{i.e.}\@\xspace}

\newcommand{\pwscf}{\textsc{Pwscf}\@\xspace}

\newcommand{\abinitio}{\textit{ab initio}\@\xspace}
\newcommand{\Abinitio}{\textit{Ab initio}\@\xspace}
\newcommand{\insitu}{\textit{in situ}\@\xspace}

\journal{Scripta Materialia}

\begin{document}

\begin{frontmatter}

\title{Oxygen - dislocation interaction in titanium from first principles}

\author[SRMP]{Nermine Chaari\fnref{Framatome}}
\author[ILM]{David Rodney}
\author[SRMP]{Emmanuel Clouet\corref{CA}}
\cortext[CA]{Corresponding author}
\ead{emmanuel.clouet@cea.fr}
\address[SRMP]{DEN-Service de Recherches de Métallurgie Physique, CEA, Université Paris-Saclay, F-91191 Gif-sur-Yvette, France}
\address[ILM]{Institut Lumière Matière, CNRS-Université Claude Bernard Lyon 1, F-69622 Villeurbanne, France}
\fntext[Framatome]{Present address: Framatome, 10 rue Juliette Récamier, F-69006 Lyon, France}

\begin{abstract}
	The interaction between screw dislocations and oxygen interstitial atoms 
	is studied with \abinitio calculations in hexagonal close-packed titanium.
	Our calculations evidence a strong repulsion when the solute atoms are located in the dislocation glide plane, 
	leading to spontaneous cross-slip, which allows the dislocation to bypass the atomic obstacle. 
	This avoidance process explains several experimental observations in titanium in presence of oxygen: 
	(1) a larger lattice friction against screw dislocation motion, 
	(2) a reduction of the dislocation glide distance in prismatic planes and 
	(3) an enhancement of cross-slip in pyramidal planes.
\end{abstract}


\end{frontmatter}

In hexagonal close-packed (hcp) titanium alloys ($\alpha$-Ti alloys), plasticity is strongly influenced 
by the interactions between $1/3\,\hkl<1-210>$ dislocations and interstitial solute atoms, in particular oxygen \cite{Tyson1967a,Akhtar1975a,Conrad1981,Naka1988,Caillard2003}.
A small oxygen content leads to an important hardening, with the plastic flow controlled by thermal activation 
instead of the almost athermal plasticity observed in pure titanium. 
Such hardening cannot be explained by a simple elastic interaction between dislocations and impurities \cite{Tyson1967a,Naka1988,Yu2015}. The most likely reason is a change of dislocation core properties because microscopic observations and \insitu experiments show long rectilinear screw dislocations in titanium with oxygen 
\cite{Naka1988,Farenc1993,Yu2015,Barkia2017}, 
an evidence that the lattice friction against the glide of screw dislocations increases in presence of oxygen atoms and may control the plasticity.

Transmission electron microscopy (TEM) also shows that dislocation motion 
is intermittent at low temperature, with long time periods where the lines are immobile, 
followed by short events where the dislocations glide in prismatic planes over a long distance \cite{Farenc1993,Farenc1995,Clouet2015}.
This motion becomes more jerky when the oxygen content increases,
with a decrease of the glide distance in prismatic planes \cite{Barkia2017}.
Dislocations also cross-slip at higher rates, mostly from prismatic to first-order pyramidal planes,
but also to basal planes for well-oriented samples \cite{Churchman1954}.
Jogs and super-jogs appear on screw dislocations because of multiple cross-slip,
finally leading to the creation of dislocation dipoles and debris \cite{Naka1982,Naka1983,Farenc1993,Yu2015,Kacher2016,Barkia2017}. 

High-resolution electron microscopy gives some information on the dislocation core structure in titanium alloys
\cite{DeCrecy1983,Neeraj2005,Yu2015}.
In particular, a narrowing of the screw core has been evidenced in presence of oxygen \cite{Yu2015}.
\Abinitio calculations based on the Density Functional Theory have also proved
a suitable approach \cite{Rodney2017}. It was shown that the ground state of the screw dislocation in pure titanium
is dissociated in a first-order pyramidal plane  
and needs to transform into a metastable configuration dissociated in a prismatic plane 
to glide easily \cite{Clouet2015}, leading to a locking-unlocking mechanism,
in agreement with the intermittent glide observed in TEM \insitu straining experiments \cite{Farenc1993,Farenc1995,Clouet2015}. Moreover, previous calculations \cite{Yu2015} have evidenced a strong short-range repulsion between the prismatic core of the screw dislocation and an oxygen atom. 
Here, we extend this work to consider all possible configurations of the screw dislocation core and
in particular the pyramidal ground state,
in order to determine how oxygen atoms affect transitions between core configurations 
and what are the potential impacts on the plasticity of Ti-O alloys.

\begin{figure*}[!bth]
	\includegraphics[width=0.99\linewidth]{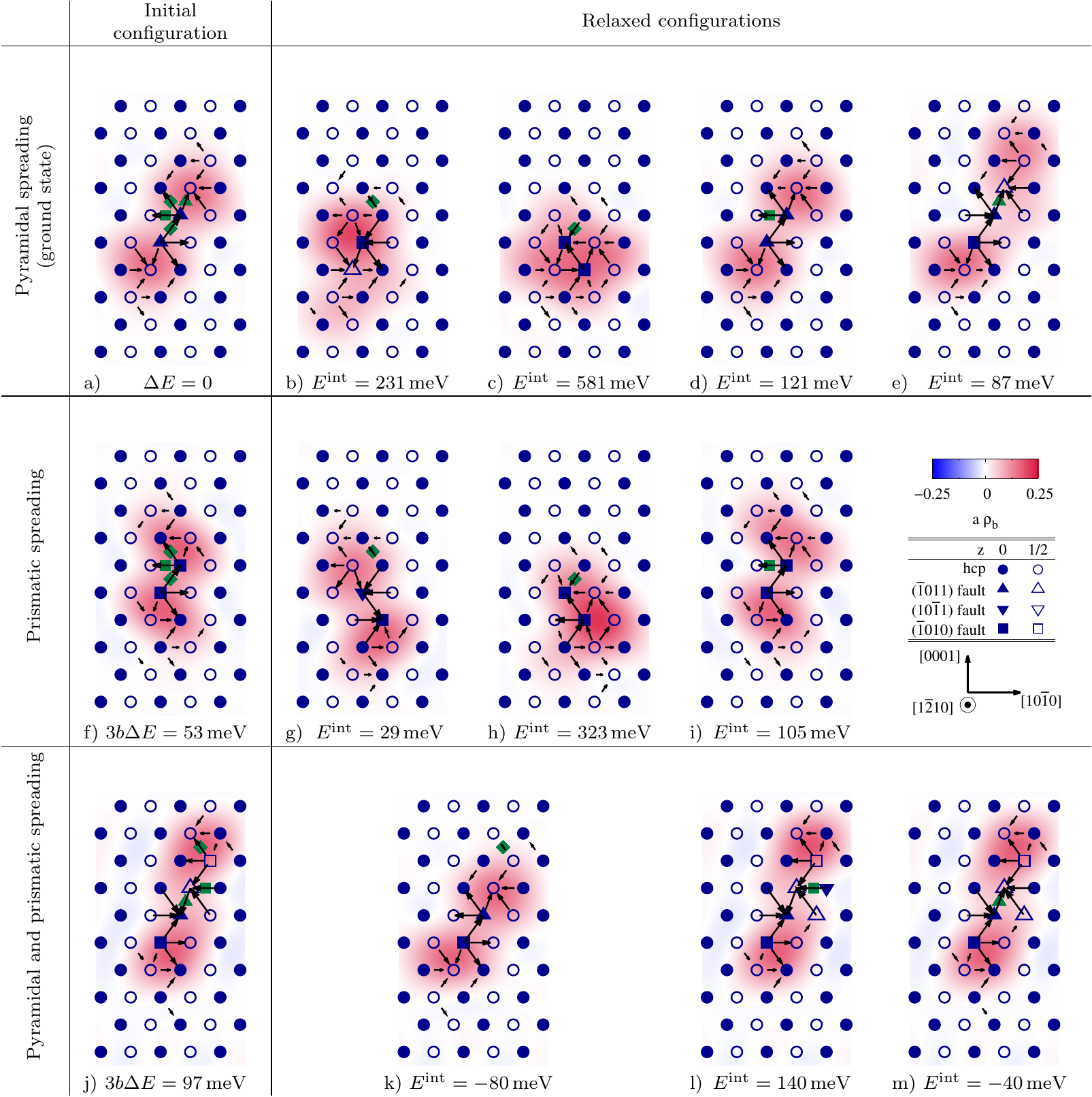}
	\caption{Core structure of a screw dislocation initially dissociated 
	(a-e) in the pyramidal plane,
	(f-i) in the prismatic plane,
	and (j-m) in both the pyramidal and prismatic planes,
	in presence of an oxygen impurity. 
	The initial configurations of the dislocation, before introduction of the oxygen atom, are shown in (a, f, j)
	with the various insertion sites for the oxygen atom shown in green.
	Relaxed structures are shown in (b, c, g, h, k) for an oxygen atom lying in an octahedral site 
	destroyed by the stacking fault ribbon of the dislocation,
	in (d, i, l) for an oxygen atom in an octahedral site created by the prismatic component of fault, 
	and in (e, m) by its pyramidal component.
	In these projections perpendicular to the dislocation line, titanium atoms are sketched by symbols
	with a color depending on their \hkl(1-210) plane in the original perfect crystal
	and with different symbols used for atoms, depending on their neighborhood in the dislocated crystal. 
	The arrows between two atomic columns are proportional to the differential displacement created
	by the dislocation in the \hkl[1-210] direction. Displacement smaller than $0.1\,b$ are not shown.
	The contour map shows the dislocation density according to the Nye tensor.
	}
	\label{fig:dislo_O}
\end{figure*}

\Abinitio calculations are performed with the \pwscf code \cite{Giannozzi2009},
in the generalized gradient approximation using the PBE functional \cite{Perdew1996}.
Ultrasoft pseudopotentials of Vanderbilt type are used 
with $3s$ and $3p$ semi-core electrons included in valence states for titanium.
Electronic wavefunctions are expanded in plane waves 
with a cutoff energy of 28\,Ry.
A regular grid of $2\times1\times7$ $k$-points is used for integration in reciprocal space
and the electronic density of states is broadened with the Methfessel-Paxton function
using a spreading of 22\,mRy.
Atoms are relaxed until all atomic force components are below 10\,meV/\AA.

To preserve periodic boundary conditions, 
we introduce a dislocation dipole in the simulation cell with a quadrupolar geometry \cite{Clouet2012,Rodney2017}.
The periodicity vectors, before introduction of the dipole, are 
$\vec{a}_x =6 a \hkl[10-10]$, $\vec{a}_y =8 c \hkl[0001]$,
and $\vec{a}_z = 3\vec{b} = a \hkl[1-210]$, corresponding to 576 lattice sites. This setup correctly predicts that the dislocation ground state is dissociated in a first-order pyramidal plane \cite{Clouet2015} (Fig. \ref{fig:dislo_O}a).
The prismatic core (Fig. \ref{fig:dislo_O}f)
has a positive relative energy $\Delta E = 5.96$\,meV/\AA. 
We also consider another metastable configuration (Fig. \ref{fig:dislo_O}j) \cite{Clouet2015},
which is spread in both prismatic and pyramidal planes 
and has a higher relative energy, $\Delta E = 11.0$\,meV/\AA.
For the oxygen atom, our calculations predict that the most stable insertion site is octahedral ($O$), 
with excess energies for the other metastable positions 
in agreement with previous \abinitio studies \cite{Wu2011}
(see Supplementary material).


To study the interaction between screw dislocations and oxygen atoms, we start from a dislocation relaxed in pure titanium in one of the three reference configurations.
We then introduce an oxygen atom in the vicinity of the dislocation core. Since the simulation cell is three Burgers vectors long in the $z$ direction, we effectively insert an oxygen atom every three sites along the dislocation line.
The interaction energy is defined as
\begin{equation}
	E^{\rm int} = E_{\rm dislo-O} - E_{\rm O} - E_{\rm dislo} + E_{\rm bulk},
	\label{eq:Eint}
\end{equation}
where $E_{\rm dislo-O}$, $E_{\rm O}$, $E_{\rm dislo}$ and $E_{\rm bulk}$
are the energies of the same supercell with respectively both the dislocation dipole and the oxygen atom,
with only the oxygen atom, only the dislocation dipole, and no defect nor solute. 
Attraction corresponds to a negative interaction energy. 
The dislocation reference energy $E_{\rm dislo}$ 
is that of the initial dislocation configuration before insertion of the oxygen atom.
As only one dislocation is decorated with an oxygen atom, part of this interaction energy 
arises from a variation of the elastic interaction between the two dislocations when the interaction 
with the oxygen atom is strong enough to induce a motion of the dislocation.
Relaxed configurations are then analyzed with the help of differential displacement maps \cite{Vitek1970} 
and the screw component of the Nye tensor \cite{Hartley2005}.
The results of our calculations are summarized in Fig. \ref{fig:dislo_O}, 
where each line corresponds to one of the three initial core configurations 
and the columns show the relaxed cores, 
without oxygen atoms in the first column and with oxygen atoms inserted in different positions in the other columns.

We first consider the case when an oxygen atom is inserted in the dislocation core in an octahedral position of the perfect crystal that is destroyed by the stacking fault ribbon when the dislocation dissociates 
\cite{Ghazisaeidi2014a,Yu2015,Kwasniak2016,Chaari2017}. These sites are shown as diamonds in Fig. \ref{fig:dislo_O} in the pyramidal and prismatic planes. These positions are strongly repulsive for all three initial core configurations, which all become unstable and relax to new dislocation cores.
The ground state, initially dissociated in a pyramidal plane (Fig. \ref{fig:dislo_O}a), relaxes to a compact core 
(Fig. \ref{fig:dislo_O}b,\,c). The same compact core is obtained with the metastable prismatic configuration (Fig. \ref{fig:dislo_O}f)
when the oxygen atom is inserted at the dislocation center (Fig. \ref{fig:dislo_O}h).
For an insertion site away from the dislocation center, 
the prismatic configuration is destabilized and falls back into the pyramidal ground state (Fig. \ref{fig:dislo_O}g).
The same destabilization is obtained with the mixed pyramidal-prismatic metastable core (Fig. \ref{fig:dislo_O}j), which also reverts back to the pyramidal core (Fig. \ref{fig:dislo_O}k).
In that case, the interaction energy is negative ($E^{\rm int}=-80$\,meV)
but this is mainly due to the energy variation associated with the change of core configuration ($3b\,\Delta E=97$\,meV).

The compact core identified in Fig. \ref{fig:dislo_O}b, c and h is of high energy and can therefore be produced only under an applied stress.
A narrowing of the pyramidal and prismatic spreading induced by oxygen atoms however agrees with the high-angle annular
dark-field scanning transmission electron microscopy (HAADF-STEM)
observations of Yu \etal \cite{Yu2015}, who reported that the screw dislocation core
is smaller in a titanium alloy containing 0.3 than 0.1\,wt.\,\% oxygen,
with oxygen atoms occupying octahedral sites in the core vicinity. 
Similar high-energy compact cores are also induced by substitutional solutes, 
in particular indium \cite{Kwasniak2017}.

The second case considered here is when the oxygen atom is introduced in insertion sites created by the stacking fault ribbon when the dislocation dissociates.
Interstitial sites $O_{\rm P}$ (square symbols in Fig. \ref{fig:dislo_O})
can be found in both the prismatic and pyramidal faults. 
Our calculations show that these insertion sites are stable and do not induce any core reconfiguration 
(Fig. \ref{fig:dislo_O}d,\,i,\,l).
The $O_{\rm P}$ sites nevertheless lead to a repulsive interaction, 
with interaction energies in the range 100\,-\,140\,meV.
These interaction energies are higher than reported by Yu \etal (50\,meV) \cite{Yu2015} 
but the difference may be because these authors used a cell with a smaller height $h=b$,
thus corresponding to the dislocation interaction with a full row of oxygen atoms.

The third case is when the stacking fault ribbon lies in the pyramidal plane and 
the oxygen atom is inserted in a new $O_{\pi}$ octahedral site created by the pyramidal fault
(triangle symbols in Fig. \ref{fig:dislo_O}).
These interstitial sites correspond to octahedral sites 
in the twined hcp crystal \cite{Chaari2017},
since the pyramidal fault is a two-layer twin in the pyramidal stacking \cite{Chaari2014,Chaari2014a}. 
Both the pyramidal and mixed pyramidal-prismatic cores relax towards the same core shown in Figs.\,\ref{fig:dislo_O}e and m,
which is close to the mixed pyramidal-prismatic core in pure titanium with the oxygen atom at the center of the core. Compared to the pyramidal core, the interaction is again repulsive ($E^{\rm int}=87$\,meV).
 
From the above calculations, we conclude that the lowest-energy configuration in Ti-O alloys remains the pyramidal core without oxygen atoms. 
All other configurations, although metastable, have a higher energy. 
Therefore, we can anticipate that in Ti-O alloys, oxygen atoms do not segregate on screw dislocations.
To glide, the pyramidal core needs to transform into a prismatic core. When the latter comes across an oxygen atom in an octahedral site, \ie a site that will be destroyed by the fault, the first site encountered by the dislocation leads to the pyramidal core in Fig. \ref{fig:dislo_O}g. The compact core in Fig. \ref{fig:dislo_O}h is less likely because of its much higher energy 
and because it corresponds to an octahedral site further inside the dislocation core.
The decorated prismatic core in Fig. \ref{fig:dislo_O}i is also unlikely,
not only because of its high energy, but also because  
it requires migration of the oxygen atom to a newly created octahedral site while the dislocation is in an excited state 
and glides in the prismatic plane. 

At low temperatures, when oxygen long-range diffusion is not active, the most likely scenario is therefore that, when the glissile prismatic core encounters an oxygen atom, it locally reverts back to a pyramidal core to avoid the oxygen atom. This scenario leads to dislocation cross-slip and is thus specific to the screw orientation, in agreement with experiments \cite{Naka1988,Farenc1993,Yu2015,Barkia2017},
which report that screw dislocations are more affected by the presence of oxygen atoms than edge dislocations. This scenario is also consistent with TEM observations \cite{Barkia2017}, which show 
that an oxygen addition destabilizes prismatic slip, 
leading to shorter glide distances in prismatic planes between pinning points. 
As the gliding dislocation locally cross-slips to bypass oxygen atoms, 
jogs are created
\cite{Yu2015,Chaari2017} in agreement with the post-mortem TEM observations of Williams \etal \cite{Williams1972}
showing that in the dislocation microstructure that develops in presence of oxygen atoms, most dislocations are jogged screw dislocations.
These jogs are responsible for an increased lattice friction 
acting against dislocation glide. 
They also lead in $\alpha$-titanium alloys to creep rates
controlled by the motion of jogged screw dislocations \cite{Viswanathan2002a,Viswanathan2002b}.
When the temperature becomes high enough to allow for oxygen diffusion, this bypass mechanism
will be in competition with oxygen migrating out of the repulsive region, 
thus allowing the dislocation to remain in its glide plane.

A reconstruction of the dislocation core into a pyramidal core induced by the oxygen atoms 
also allows to understand why cross-slip in pyramidal planes becomes more active
with an increase of the oxygen content \cite{Yu2015,Barkia2017}. 
Indeed, once stabilized, the pyramidal core can glide in its pyramidal habit plane by nucleating 
kink pairs \cite{Clouet2015}.

Comparing the present results  
with Ref. \cite{Chaari2017} on the interaction of screw dislocations with oxygen atoms in zirconium, 
we see that the same repulsive interaction resulting from the destruction of oxygen insertion sites 
by the stacking fault ribbon and leading to a change of dislocation core configuration
is obtained in zirconium and titanium when the oxygen atom lies in its natural octahedral insertion sites. 
Therefore, in both metals, the screw dislocations gliding in prismatic planes undergo a stronger lattice friction
because of the creation of jogs allowing to bypass the solute atoms.
Moreover, cross-slip in pyramidal planes is enhanced in both metals by oxygen atoms.
However, while in titanium, cross-slip in pyramidal  planes is already active at low temperatures 
even for a very low oxygen content \cite{Clouet2015,Caillard2018},
in zirconium, plasticity is confined to prismatic planes at low temperatures 
and cross-slip appears only at higher temperatures
thanks to thermal activation or interaction with oxygen atoms \cite{Chaari2017}. 
The origin of the lattice friction in the principal slip system also differs between both metals:
in zirconium, this friction is purely extrinsic and only arises from the interaction 
with impurities, in particular oxygen, 
whereas an intrinsic friction still remains in pure titanium at low temperature \cite{Biget1989,Clouet2015,Caillard2018}.

\vspace{0.5cm}
\linespread{1}
\small

\textbf{Acknowledgements} -
This work was performed using HPC resources from GENCI-CINES and -TGCC (Grants 2017-096847).
The authors also acknowledge PRACE for access to the Curie resources based in France at TGCC 
(project PlasTitZir).
DR acknowledges support from LABEX iMUST (ANR-10-LABX-0064) of Université de Lyon 
(program ``Investissements d'Avenir'', ANR-11-IDEX-0007)

\section*{References}
\bibliographystyle{model1a-num-names}
\bibliography{chaari2018}

\end{document}


\begin{frontmatter}

\title{Oxygen - dislocation interaction in titanium from first principles \\
Supplementary material}

\author[SRMP]{Nermine Chaari\fnref{Framatome}}
\author[ILM]{David Rodney}
\author[SRMP]{Emmanuel Clouet\corref{CA}}
\cortext[CA]{Corresponding author}
\ead{emmanuel.clouet@cea.fr}
\address[SRMP]{DEN-Service de Recherches de Métallurgie Physique, CEA, Université Paris-Saclay, F-91191 Gif-sur-Yvette, France}
\address[ILM]{Institut Lumière Matière, CNRS-Université Claude Bernard Lyon 1, F-69622 Villeurbanne, France}
\fntext[Framatome]{Present address: Framatome, 10 rue Juliette Récamier, F-69006 Lyon, France}

\end{frontmatter}

\begin{table}[!tb]
\begin{minipage}{\linewidth}
\renewcommand{\footnoterule}{}	
	\caption{Energy of an O atom in a perfect hcp lattice for different configurations:
	octahedral ($O$) taken as the reference, basal tetrahedral ($BT$) 
	and crowdion ($C$). 
	Results are given for different supercells 
	defined by their number $N_{\rm sites}$ of lattice sites
	The supercell is based either on the hcp primitive cell 
	or corresponds to the cell used for dislocation modeling.}
	\label{tab:energy_O}
	\centering
	\begin{tabular}{ccccc}
		\hline
		\multirow{2}{*}{cell}	& \multirow{2}{*}{$N_{\rm sites}$}	& \multicolumn{2}{c}{$\Delta E$ (eV)} \\
		\cline{3-5}
								&	& $O$	& $BT$	& $C$ \\
		\hline
		$4\times4\times3$\footref{fn:Wu2011} hcp	& 96	& 0.	& 1.19	& 1.88	\\
		$4\times4\times3$\footref{fn:present} hcp	& 96	& 0.	& 1.16	& 1.84	\\
		$5\times5\times4$\footref{fn:present} hcp	& 200	& 0.	& 1.13	& 1.83	\\
		$6\times8\times3$\footref{fn:present} dislo	& 576	& 0.	& 1.20	& ---  \\
		\hline
	\end{tabular}
	\footnotetext[1]{\label{fn:Wu2011} Ref. \cite{Wu2011}}
	\footnotetext[2]{\label{fn:present} Present work}
\end{minipage}
\end{table}

\section*{References}
\bibliographystyle{model1-num-names}
\bibliography{chaari2018}